# Intracavity-birefringence-enabled soliton states and wavelength control in the (C + L)-band fiber lasers

Chuangkai Li, Feng Ye, Hong Jin, Xuanyi Liu, H. Y. Fu, Boris A. Malomed, and Qian Li

*Abstract*—We demonstrate a compact all-fiber nonlinear-polarization-evolution (NPE) fiber laser capable of outputting wavelength-manipulated multiple laser states in the C + L band. Leveraging the intracavity birefringence-induced filtering effect, without external spectral filters, the laser achieves wavelength-tunable conventional solitons (CSs) and soliton molecules (SMs) with a tuning range of 72.85 nm and 45.54 nm, respectively, by adjusting the polarization states and pump power. Furthermore, wavelength-switchable harmonic mode-locking (over 1 GHz) and dual-wavelength mode-locking can also be achieved in the C + L band. To the best of our knowledge, this is the first experimental realization of wavelength-manipulated multiple soliton states covering the C + L band in a single-fiber-laser oscillator. Two examples of digital multi-letter encoding and transmission are further demonstrated on this platform by means of pump-driven wavelength/soliton states switching. By enabling controllable multi-state solitons and broad spectral tunability within a compact architecture, our work provides a promising source for spectrally diverse and temporally stable pulsed light sources that hold significant promise for applications in microscopy, bioimaging, and LiDAR.

## I. INTRODUCTION

ULTRAFAST fiber lasers are extensively utilized in industrial and scientific applications, such as optical frequency comb [1], micromachining [2], ultrafast physics[3], etc., due to their compactness, sturdiness, reduced thermal effects, and low construction cost [4]. The operation wavelength is one of the crucially important parameters of the lasers. Endowing ultrafast solitons with wavelength-tunable characteristics can significantly enhance their applications in microscopy, bio-imaging, LiDAR, etc., as well as in fundamental studies[5], [6], [7]. Although the gain-carrying fiber inherently provides a broad emission window, the practically accessible tuning range of the wavelength-tunable fiber lasers (WTFLs) is not totally determined by the gain bandwidth itself, but rather by the gain competition and the gain/loss distribution band. To overcome this limitation, intracavity filters are employed to facilitate the wavelength selection and stabilization, thereby allowing broader and more flexible tuning within the entire gain bandwidth [8], [9], [10]. The intracavity filters can be broadly classified into two categories: the physical filters and artificial ones. Physical filters, such as gratings [11], [12], [13] and long-period fiber gratings[14], [15], enable center-wavelength selection through intrinsic spectral filtering in the cavity. While WTFLs based on physical filters exhibit excellent reproducibility, their reliance on discrete high-precision components inevitably increases the structural complexity of the laser cavity and poses challenges for the system's integration and scalability. On the other hand, artificial filters, such Lyot [16], [17], [18], [19], Sagnac [20], [21], [22], [23], and multimode interference filters [24], [25], as well as variable optical attenuators (VOAs) [9], [10], [26], enable the control of the gain/loss profile and provide a wavelength-tunable output through filtering effects induced by specialized fibers or structures. These solutions have garnered significant interest due to their cost effectiveness. However, the use of specialized fibers and structures inevitably complicates cavity architecture.

Intracavity birefringence-induced filtering (IBIF) effect offers an excellent approach to the design of artificial filtering, that leverages the filtering effects caused by the birefringence in the laser cavity. In comparison to the above-mentioned filtering techniques, IBIF effect removes the need of additional filtering devices, specialized fibers, or complex structures. The components required for IBIF effect is similar to the nonlinear-polarization-evolution (NPE) mode-locking mechanism, indicating that the same device can simultaneously achieve mode-locking and filtering functionalities. Consequently, wavelength-tunable fiber lasers using this technology feature a compact architecture. The recent work on erbium-doped fiber (EDF) lasers has helped to achieve wavelength-controlled operations with multiple soliton states, including dissipative solitons (DSs) [27], [28], conventional solitons (CSs) [29], [30], stretched pulses (SP) [31], harmonic mode-locking (HML) [32], etc. These independent studies have demonstrated the effectiveness of constructing WTFL via IBIF effect, emphasizing its compact

This work was supported by Shenzhen International Cooperation Research Project (GJHZ20220913144206012), Guangdong Provincial Key Laboratory of In-Memory Computing Chips (2024B1212020002). *(Chuangkai Li and Feng Ye contributed equally to this work.) (Corresponding authors: Qian Li).*
Chuangkai Li, Feng Ye, Hong Jin and Qian Li are with School of Electronic and Computer Engineering, Peking University, Shenzhen 518055, China (e-mail: 2201212765@stu.pku.edu.cn; ye-feng@pku.edu.cn; jinh@stu.pku.edu.cn; liqian.sz@pku.edu.cn).
Xuanyi Liu and H. Y. Fu are with Tsinghua Shenzhen International Graduate School, Tsinghua University, Shenzhen 518055, China (e-mail: liuxy728@mail.sysu.edu.cn, hyfu@sz.tsinghua.edu.cn).
Boris A. Malomed is with the Instituto de Alta Investigación, Universidad de Tarapacá, Casilla 7D Arica, Chile (e-mail: malomed@tauex.tau.ac.il).

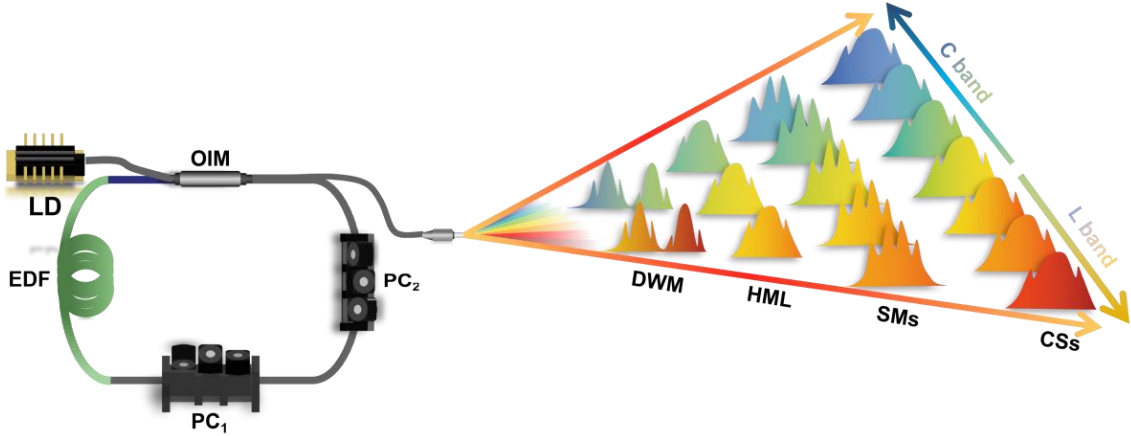

Fig. 1. The experimental setup of the wavelength-manipulated fiber laser includes the following components: OIM: optical integrated module, PC: polarization controller, EDF: erbium-doped fiber, LD: laser diode. The outputs produced by the setup include the following regimes: DWM: dual-wavelength mode-locking, HML: harmonic mode-locking, SMs: soliton molecules, and CSs: conventional solitons.

structure. While previous studies predominantly focused on the wavelength control for individual solitons, the capability of producing multiple soliton states with selectable wavelengths enhances the potential applications of fiber lasers across various fields. As an illustration, it provides an effective research platform for the investigation of switching and evolution between different soliton types and multiple wavelengths [33]. Additionally, the variation in output characteristics enriches the information-carrying capacity for optical encoding [34], [35]. Despite extensive studies of IBIF effect and NPE over the years, no reports to date have demonstrated the capability to manipulate the wavelength of more than two distinct soliton states simultaneously in the C + L band. Concurrent wavelength manipulations of multiple soliton states in an IBIF-based fiber laser remain an open challenge and await further exploration.

In this study, we present a compact intracavity wavelength-manipulation scheme, which is directly driven by the IBIF effect in the NPE mode-locking fiber-laser oscillations without the need for additional precision components. Through the adjustments of the polarization states and pump power, wide-range wavelength-tunable outputs in the form of CSs and soliton molecules (SMs) can be produced with tuning ranges of 72.85 nm and 45.54 nm, respectively, belonging to the C + L band. The wavelength switching and coexistence of the HML and dual-wavelength mode-locking (DWM) states can also be achieved. The HML refers to a regime in which pulse splitting, induced by the peak-power clamping effect, leads to the formation of multiple pulses per round trip, thereby increasing the repetition rate to values exceeding 1 GHz in our case. Leveraging this wavelength-control laser system, two cases of multi-letter encoding are demonstrated through power-driven transitions between SMs at 1570.30 nm and CSs at 1601.20 nm, as well as between multiple solitons (MSs) at 1571.92 nm and CSs at 1599.50 nm. This work presents the first demonstration of multiple soliton states with wavelength-manipulated performance in an IBIF-based fiber laser covering the C + L band, offering a new paradigm in soliton dynamics.

II. EXPERIMENTAL SETUP AND THEORETICAL ANALYSIS

The configuration of the proposed all-fiber wavelength-manipulated fiber laser is plotted in Fig. 1. An optical integrated module (OIM), which consists of a wavelength-division multiplexer (WDM), a polarization-sensitive isolator (PS-ISO), and a 90:10 optical coupler (OC), is used to make the laser structure more compact, and the details are available in our previous research [36], [37]. Two polarization controllers (PCs) are utilized to control the polarization states and develop the NPE mode-locking mechanism in conjunction with the PS-ISO. An 88-cm-long section of the erbium-doped fiber (EDF80, OFS, $\beta_2 = 0.0612$ ps$^2$/m) is inserted in the cavity as the gain medium, pumped by a commercial 980 nm laser diode (LD) with the maximum power of 1.2 W. The other fibers used in the laser cavity include a 0.91-m Hi1060 and a 3.70-m SMF-28e, yielding a total cavity length of ~5.50 m and net dispersion of -0.025 ps$^2$.

The wavelength-manipulated capability of the fiber laser is attributed to the IBIF effect, whose transmission function **can** be expressed as the Jones matrix [38], [39]:

$$T = \sin^2(\theta_1)\sin^2(\theta_2) + \cos^2(\theta_1)\cos^2(\theta_2) \\ + 0.5\sin(2\theta_1)\sin(2\theta_2)\cos(\Delta\varphi_L + \Delta\varphi_{NL}) \quad (1)$$

where $\theta_1$ and $\theta_2$ are the azimuthal angles between the polarization direction of the input light, polarizer and the fiber's fast axis, respectively. Further, $\Delta\varphi_L$ and $\Delta\varphi_{NL}$ are the linear and nonlinear phase delay, respectively, denoted as

$$\Delta\varphi_L = \Delta\varphi_0 + 2\pi(1 - \partial\lambda/\lambda_S)L/L_b \quad (2)$$

$$\Delta\varphi_{NL} = 2\gamma LP\cos(2\theta_1)/3 \quad (3)$$

where $L$, $L_b$, $P$, $\delta\lambda$, and $\lambda_S$ represent the length of the birefringent fiber, birefringence beat length, instantaneous power of the input pulse, wavelength detuning, and central wavelength, respectively. By precisely adjusting PCs, the values of $\theta_1$, $\theta_2$ and $B_m$ can be optimized, enabling the control of the modulation depth, spectral bandwidth, and filtering position, respectively.

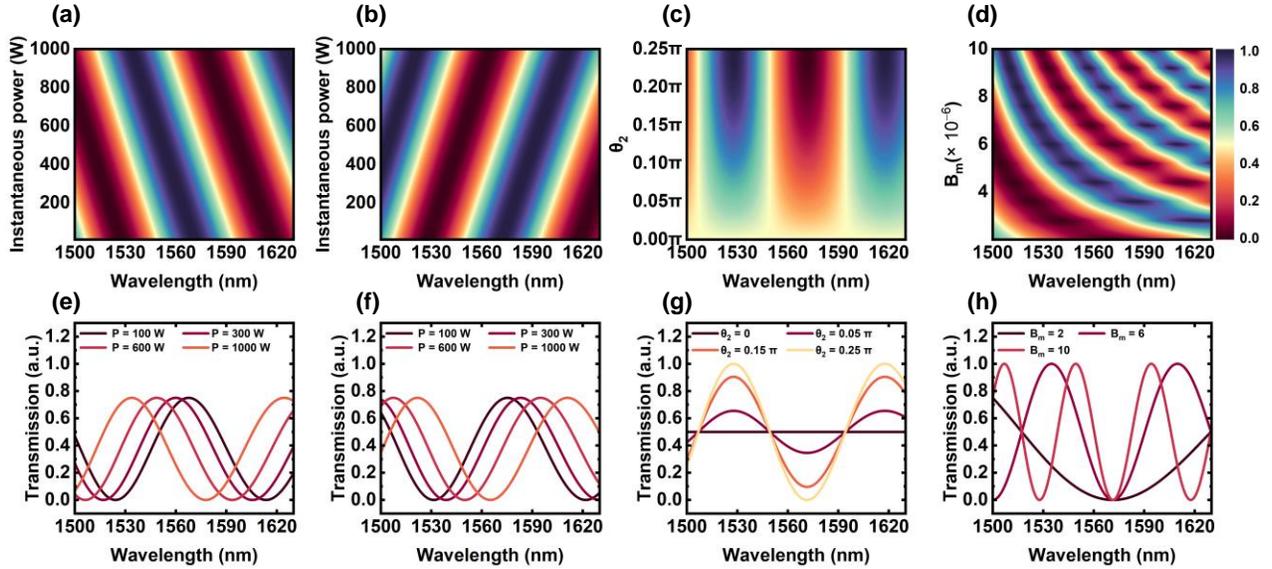

Fig. 2. The simulated spectral transmission curves as functions of $P$ ((a), (b), (e), (f)), $\theta_2$ ((c), (g)), and $B_m$ ((d), (h)). With $\theta_1$ in the ranges ($\pi/4$, $\pi/2$) and (0, $\pi/4$), increasing peak power results in a blue shift and a red shift in the peak position of the filtering curve, respectively. By keeping other parameters constant, adjustments to $\theta_2$ and $B_m$ enable control over the modulation depth and bandwidth of the filtering curve. The parameters used for the spectral transmission curve mentioned above are summarized as follows: (a), (e) $\theta_0 = 0$, $\theta_1 = \pi/3$, $\theta_2 = 5\pi/6$, $B_m = 5\times10^{-6}$; (b), (f) $\theta_0 = 0$, $\theta_1 = \pi/6$, $\theta_2 = 2\pi/3$, $B_m = 5\times10^{-6}$; (c), (g) $\theta_0 = 0$, $\theta_1 = \pi/4$, $P = 500$ W, $B_m = 5\times10^{-6}$; (d), (h) $\theta_0 = 0$, $\theta_1 = \pi/4$, $\theta_2 = 3\pi/4$, $P = 500$ W.

The IBIF simulated transmission curve, calculated by (1)-(3), is displayed in Fig. 2, illustrating the tunable filtering characteristics. This tunability is governed by the PCs, which adjust the azimuthal angles between the polarization direction of the input light, polarizer, and the fast axis of the fiber $\theta_1$, $\theta_2$, and the birefringence coefficient $B_m$ to regulate the modulation depth, spectral bandwidth, and gain peak. Fig. 2(a), (b), (e), and (f) demonstrate the spectral evolution following the variation of $P$. As depicted in Fig. 2(a), (e) and Fig. 2(e), (f), when $\theta_1$ belongs to intervals ($\pi/4$, $\pi/2$) and (0, $\pi/4$), the transmission spectra move to the C-band (L-band) for negative (positive) $\Delta\varphi_{NL}$ with the increase of $P$. The controlled gain-peak displacement enables the pump-driven wavelength switching in the laser system, which has been verified in previous studies [40]. For the values of $\theta_1$, $P$, and $B_m$ fixed as $\pi/6$, 500 W, and $5\times10^{-6}$, the modulation depth becomes directly controllable through the adjustment of $\theta_2$. As shown in Fig. 2(c), (g), the progressive increase of $\theta_2$ from 0 to $\pi/4$ leads to the linear enhancement of the modulation depth from 0 to 1. The free spectral range (FSR) of the transmission curve exhibits the inverse proportionality to $B_m$, as evidenced by Fig. 2(d), (h). Increasing $B_m$ from $2\times10^{-6}$ to $10\times10^{-6}$ produces a simultaneous reduction in both the transmission bandwidth and FSR. Previous research has confirmed that, adjusting these parameters, it is possible to modify the linear and nonlinear phase shifts, thus achieving wavelength tuning [41], [42]. This IBIF-based method eliminates the requirement for external filters, enabling the direct output of the WTFL by controlling the IBIF-based transmission curve in the laser cavity. Similar to the conventional birefringent filtering techniques [43], the introduction of a phase difference is at the core of the birefringence-induced filtering. After passing the polarizer, the light beam becomes linearly polarized and propagates along the fiber's fast and slow axes. Nonlinear phase shifts, induced by the fiber, and linear phase shifts, caused by squeezing the fiber by the PC, emerge during the transmission through the laser cavity. Due to the phase difference, the beams interfere at the polarizer, resulting in the suppression of certain spectral components and creating a filtering effect. Therefore, controlling the PCs and pump power, it is possible to control the linear and nonlinear phase shifts and adjust the filtering transmission curve.

III. EXPERIMENTAL RESULTS AND DISCUSSIONS

A. Wavelength tuning of CSs and SMs in the C + L band

Assisted by the IBIF effect, the wavelength tuning of the proposed compact all-fiber laser can be achieved by adjusting the PCs and pump power in the laser cavity. As shown in Fig. 3(a), the central CSs wavelength can be tuned from 1531.62 nm to 1604.47 nm, covering the C + L band with a tuning range of 72.85 nm. The CSs spectra with the Kelly sidebands confirm the operation in the anomalous-dispersion regime. The mode-locking states in the L-band coexist with continuous-wave (CW) light at a wavelength of ~1570 nm that can not be eliminated. This might be attributed to the strong mode competition in the homogeneously broadened gain medium [9]. The corresponding 3-dB spectral bandwidths at different wavelengths ranging from 3.70 nm to 15.93 nm are summarized in Fig. 3(c). The filtering bandwidth of the IBIF curve is inversely proportional to the cavity length. While the WTFL cavity facilitates a broader spectral output, it also simultaneously increases the susceptibility to the intracavity mode competition, which can moderately disrupt the continuity of the wavelength tuning. In this work, mode-locked states near 1593 nm and 1530 nm exhibit narrower 3-dB spectral bandwidths. For instance, the mode locking that operates at wavelengths 1531.62 nm and 1593.06 nm demonstrates the spectral full width at half-maximum (FWHM) of 3.70 nm and 7.58 nm, respectively. In contrast, the mode-locked states outside of these ranges exhibit spectral bandwidths exceeding 10 nm. Further optimization of the intracavity parameters can address this issue, which has been demonstrated in the previous study [37]. The fundamental repetition rates of the CSs at different wavelengths were measured by a radio-frequency (RF) signal analyzer. The corresponding RF spectra, with a resolution bandwidth (RBW) of 1 Hz, are presented in Fig. 3(b) and recorded in Fig. 3(d). As the mode-locking central wavelength shifts from the C-band to the L-band, the fundamental repetition rate gradually decreases from 37393480 Hz to 37390940 Hz, with a frequency difference of 2540 Hz. This behavior is attributed to different group velocities of optical pulses at varying wavelengths in the cavity.

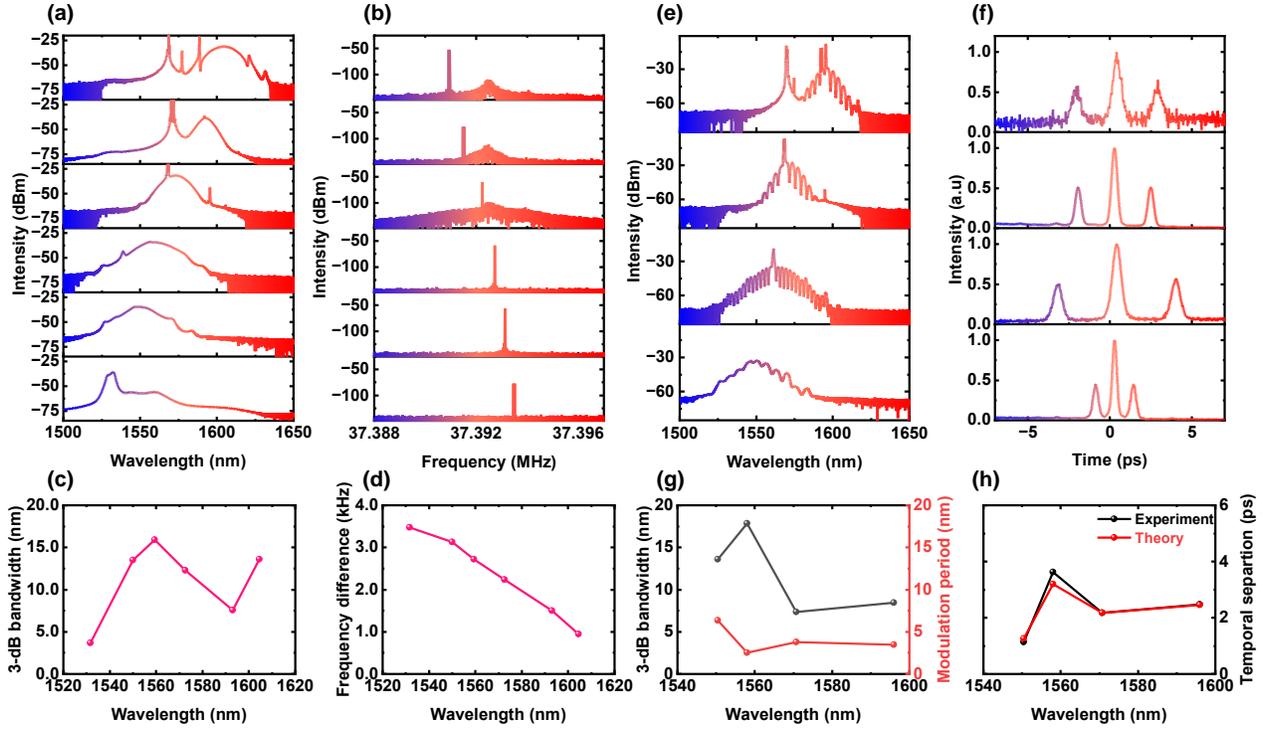

Fig. 3. Output characteristics of the wavelength-manipulated CSs and SMs. (a) Optical spectra; (b) RF spectra; (c) the 3-dB bandwidth; (d) the frequency difference (normalized to 37.39 MHz) of CSs from 1531.62 nm to 1604.47 nm; (e) optical spectra; (f) AC traces; (g) the 3-dB bandwidth and modulation period; (h) the experimental and theoretical values of the temporal size of SMs (the separation between the coupled solitons) in the wavelength interval 1550.46 nm to 1596.00 nm.

By further adjusting the intracavity gain support and PCs, SMs operating at different wavelengths can be generated from 1550.46 nm to 1596.00 nm with a tuning range of 45.54 nm. Distinct periodic modulations, observed in the spectra shown in Fig. 3(e) serve as the characteristic signature of SMs. The AC traces in Fig. 3(h) exhibit three peaks with an intensity ratio of 1:2:1, further validating the formation of the dual-soliton SMs composed of two solitons with identical energy. The relationship between the spectral modulation period and the temporal separation can be expressed as $\Delta\lambda = (\lambda_c^2)/(c\Delta\tau)$ [44], where $\Delta\lambda$, $\lambda_c$, $c$, and $\Delta\tau$ represent the spectrum modulation period, the spectrum central wavelength, the light speed in the vacuum, and the temporal separation of the SMs, respectively. The corresponding 3-dB bandwidth, spectral modulation period, experimental and theoretical separations at different wavelengths are summarized in Fig. 3(g), (h). The SMs temporal size (separation between the coupled solitons) remains ~4 ps across the entire wavelength tuning range. The experimental values of the separation align well with the theoretical values. By fine-tuning the pump power and polarization states, the laser retains the ability to generate various configurations of SMs. In the Supplementary Material, Tables S1 and S2 summarize representative operating conditions for CSs and SMs at different central wavelengths in the tuning range. These datasets clearly and comprehensively reveal the practical operating conditions for CSs and SMs generation at various wavelengths.

*B. Wavelength-switchable harmonic and dual-wavelength mode-locking*

The wavelength-tuning performance of the proposed laser enables the mode-locked states in the laser cavity to access more wavelength options. We demonstrate wavelength switching operations for two soliton states, *viz*., HML and DWM, in the C +

L band. The output characteristics with the highest harmonic repetition rate are depicted in Fig. 4, and additional HML information for other harmonic orders is elaborated in Fig. S1 and Table S3 of the supporting information.

We first utilized the IBIF effect to adjust the emission of CW light by the laser near the mode-locking region. By controlling the pump power and PCs, HML and DWM at different central wavelengths could be realized. Figure 4(a)-(i) illustrates the output characteristics of HML operating at wavelengths of 1530.57 nm, 1567.75 nm, and 1599.98 nm with 3-dB bandwidths of 5.11 nm, 0.88 nm, and 12.11 nm, respectively. The HML operating at 1599.98 nm shows more pronounced Kelly sidebands and strong CW signal cluster near 1570 nm compared to the other two regions at shorter wavelengths, resembling the optical spectrum previously observed under fundamental mode-locking in Fig. 3. The time-domain pulse signals for the three different wavelength regions are displayed in Fig. 4(b), (e), (h), with time gaps of 0.70 ns, 0.70 ns, and 2.23 ns, respectively. RF spectra reveal the repetition rates located at 1.42092 GHz, 1.42090 GHz, and 448.69 MHz, corresponding to the 38th, 38th, and 12th HML orders, respectively. To quantify the temporal stability of the pulse train in HML, the supermode suppression level (SSL) is defined as the power difference between the harmonic RF peak and the highest sideband in the RF spectrum. Generally, an SSL exceeding 20 dB is regarded as an indicator of stable HML operation. Fig. 4(c), (f), and (i) illustrate the SSL at different central wavelengths, all exceeding 20 dB, confirming the stability of the HML regimes. In our laser, the transition from fundamental mode-locking to high-order HML as the pump power increases is mainly governed by the interplay between NPE-based intensity-dependent loss and gain saturation, while the IBIF-induced spectral filtering assists in maintaining the spectral profile of the pulses.

When sufficient gain is injected into the cavity, the mode-locking can readily emerge from CW clusters, as shown in Fig. 3. By precisely controlling the PCs and gain supply in the laser cavity, DWM can be generated in both the C- and L-band. As shown in Fig. 4(j), (m), the central wavelengths of DWM in the

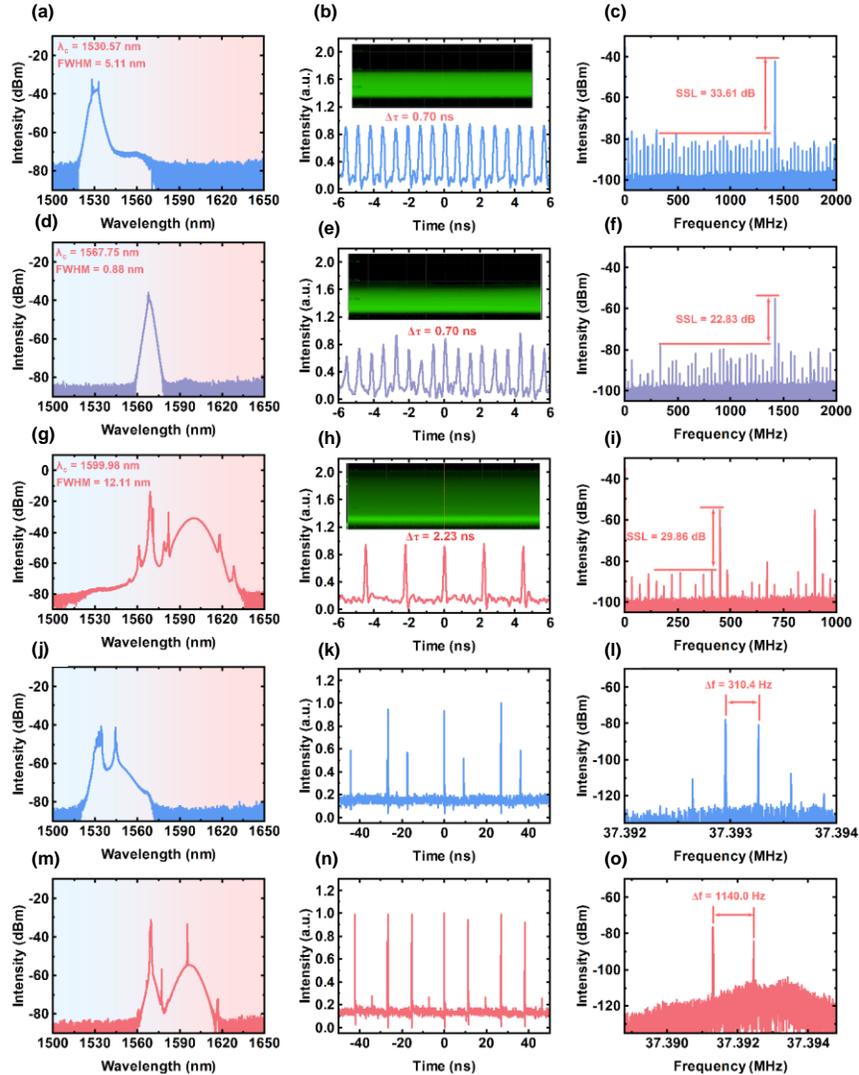

Fig. 4. The output characteristics of the HML and DWM operation in C- and L-band. (a), (d), (g), (j), (m) The output spectra. (b), (e), (h), (k), (n) the oscilloscope trains. (c), (f), (i), (l), (o) the RF spectra.

C- and L-band are ~1532 nm/~1550 nm and ~1570 nm/~1597 nm, respectively, corresponding to the spacing of ~18 nm and ~27 nm. In Fig. 4(k), (n), oscilloscope traces present two independent pulse sequences, providing direct evidence of the DWM operation regime. The RF spectra shown in Fig. 4(l), (o) reveal the frequency differences of 310.4 Hz and 1140.0 Hz in the C- and L-band, respectively, further supporting the DWM state through the emergence of two distinct frequency peaks. Moreover, we have successfully achieved the coexistence of HML and DWM. Namely, HML pulses operating at ~1567 nm and fundamental mode-locking pulses at ~1600 nm can be concurrently generated (see details in Fig. S2-4 of Supporting Information). This is also the first demonstration of the GHz-level harmonic-fundamental asynchronous pulse output.

*C. Active wavelength switching for multi-letter encoding*

In the preceding sections, we have demonstrated the existence of multiple soliton states in the laser cavity with wavelength-manipulated characteristics. Building upon the proposed light source, two examples of multi-letter encodings based on power-driven laser states and wavelength switching are developed.

We select two distinct regions where the switching of wavelengths and soliton states can be controlled solely by varying the pump power, including CSs/SMs and CSs/MSs. Figure 5 showcases the output characteristics and conversion operation performance of CSs/SMs. CSs operating at 1601.20 nm could be achieved with a pump power of 85 mW and a 3-dB bandwidth of 13.10 nm. The spectrum displays clear Kelly sidebands, confirming the operation in the anomalous-dispersion regime. The AC trace is shown in Fig. 5(b), for a duration of 397 fs with an adopted $sech^2$ pulse profile. The time-domain pulse train in Fig. 5(c) indicates the single-pulse operation regime. When the pump power is changed to 225 mW, CSs at 1601.20 nm immediately switch to SMs at 1570.30 nm. The spectrum of SMs in Fig. 5(d) exhibits regular modulation, and the corresponding AC trace in Fig. 5(e) displays three peaks with a pulse separation of 525 fs. In the time domain, upon switching the gain supply, a single pulse splits into multiple ones and eventually forms a pulse cluster. This does not affect our ability to demonstrate the switching between CSs and SMs. Reducing the pump power may lead to the conversion of a pulse cluster into single SMs. Single-pulse SMs are not stable and can easily transition into CSs. The switching process is shown in supplementary Video 2, and the corresponding oscilloscope train is presented in Fig. 5(f). A clear distinction in the pulse intensity between the two states is observed, especially when compared to the switching between CSs and MSs in Fig. S5, Supporting Information. Five consecutive state transitions were performed to test the stability of the switching process, with each switched state persisting for more than 15 minutes. Fig. 5(g) presents the output power and spectral-evolution results, with no difference before and after the transition, suggesting the strong stability of the switching between CSs and SMs. Another pump-driven wavelength switching of CSs at 1599.50 nm and MSs at 1571.92 nm is provided in Fig. S5 of the supporting information.

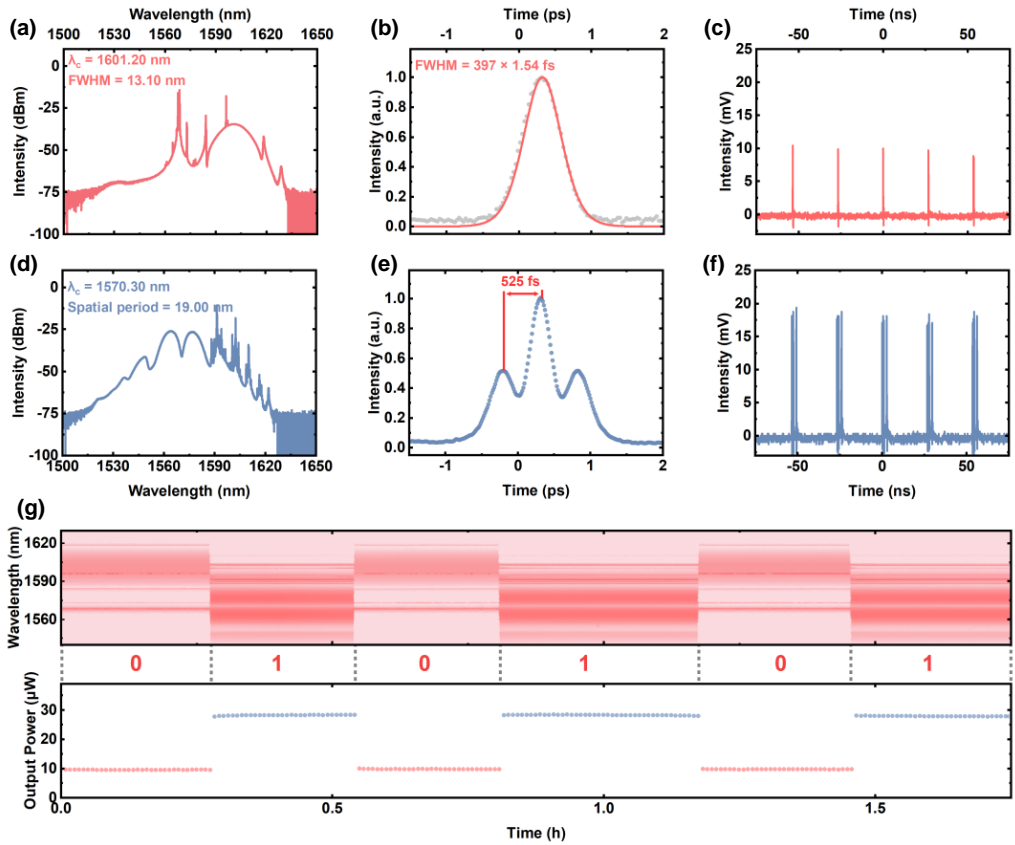

Fig. 5. Active pump-driven switching of CSs/SMs for multi-letter encoding. (a), (d) Optical spectra; (b), (e) autocorrelation traces; (c), (f) pulse trains of CSs and SMs; (g) the output spectra and power evolution for five continuous conversion operation.

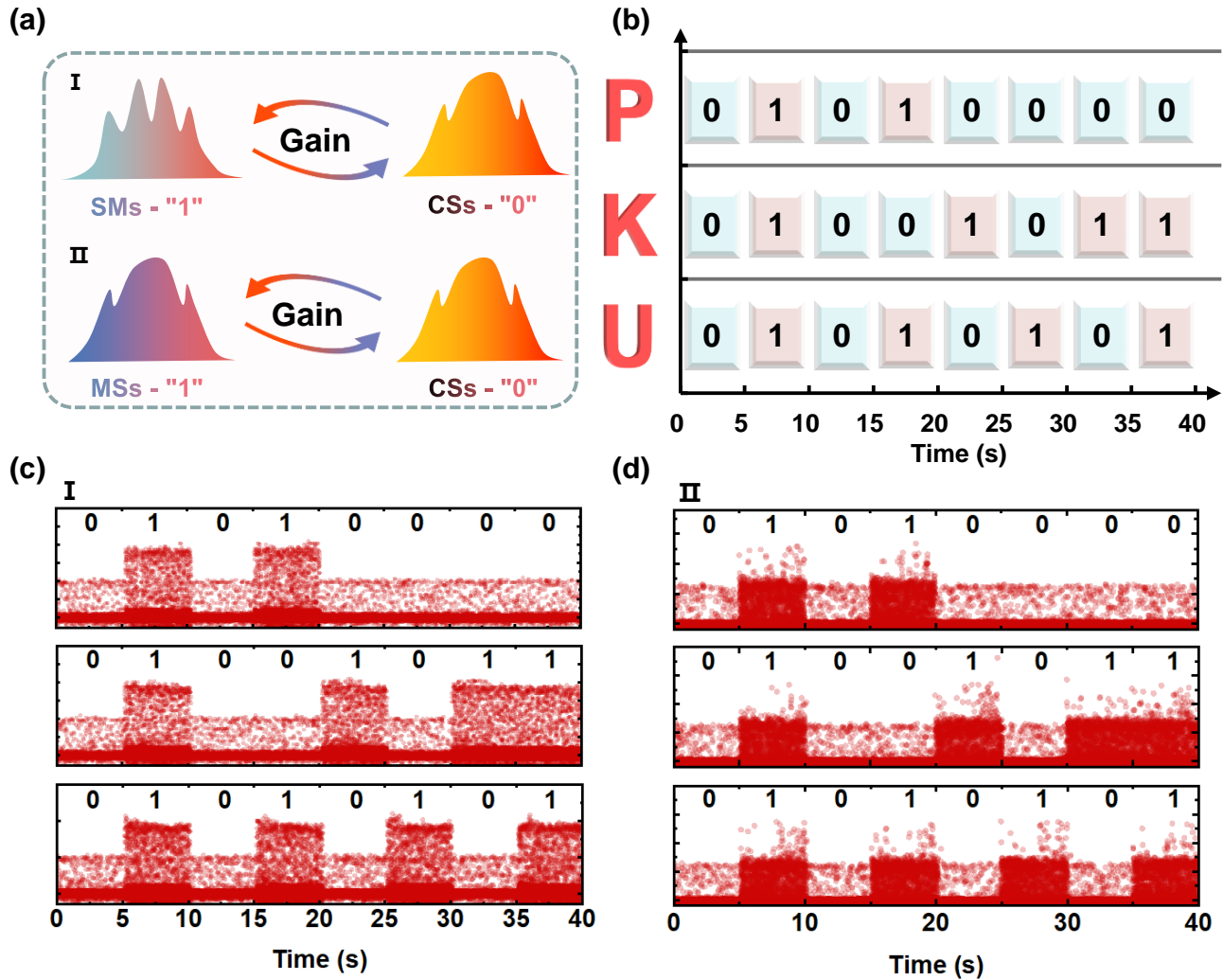

Fig. 6. The multi-letter encoding based on two soliton-switching schemes: CSs/SMs and CSs/MSs. (a) Schematic diagrams of the wavelength/state switching, based on the gain modulation; (b) The visualization of each letter. (c) For scheme I, a recorded bit stream, where CSs represents "0" and SMs represent "1". (d) For scheme II, a recorded bit stream, where CSs represents "0" and MSs represent "1".

The stable wavelength and state switching realized by the laser provides a versatile platform for optical multi-letter encoding. Two distinct encoding schemes are implemented, as summarized in Fig. 6(a). The multi-letter stream is visualized in Fig. 6(b). Using ASCII encoding, 'PKU' can be represented using eight-bit binary (P: 0101 0000; K: 0100 1011; U: 0101 0101). In Scheme I, CSs are assigned to the logical symbol "0", whereas SMs are assigned to the logical symbol "1". The two states are discriminated by their pulse amplitude. SMs exhibit a substantially higher peak power than CSs, so an intensity threshold enables robust symbol recognition. The corresponding binary stream and temporal traces for encoding the word "PKU" are shown in Fig. 6(c). In Scheme II, CSs represent the logical symbol "0", while MSs represent the logical symbol "1". In this case, the symbols are distinguished by their pulse density within each bit slot. MSs contain multiple pulses, whereas CSs contain a single pulse, allowing detection by pulse counting or energy integration, as illustrated in Figs. 6(b) and 6(d). In both schemes, the binary representation of "PKU" (P: 0101 0000; K: 0100 1011; U: 0101 0101) is generated solely by modulating the pump power, which demonstrates that pump-controlled soliton-state switching can directly support reconfigurable multi-letter optical encoding in a single all-fiber platform.

In this experiment, the pump power was manually controlled. By incorporating an electrically PC and a pump controller in the cavity, high-speed switching between multiple states and wavelengths could be achieved. This multi-letter test demonstrates the feasibility of the wavelength/states switching-based digital encoding on this platform and prospects for high-capacity all-optical storage.

## IV. Conclusion

We have demonstrated a compact wavelength-manipulated mode-locked fiber laser based on the IBIF effect, covering the C + L band. The proposed wavelength-controlled fiber laser creates mode-locked pulses of both the CSs and SMs types with the tuning ranges of 72.85 nm and 45.54 nm, respectively. HML (even over 1 GHz) and DWM in the C + L band can be realized with specific wavelengths. To the best of our knowledge, this is the first demonstration of widely wavelength-tuning operations for multiple soliton states in the C + L band. Based on the proposed wavelength-controlled mode-locked fiber laser, we have implemented two examples of the multi-letter encoding cases driven by the power-induced wavelength/ state conversion. This compact, all-fiber, widely wavelength-tunable fiber laser holds significant potential for applications to microscopy, bio-imaging, LiDAR, etc.